# High-resolution Millimeter Imaging of Submillimeter Galaxies[1]

L.J.Tacconi[(1)], R.Neri[(2)], S.C.Chapman[(3)], R.Genzel[(1,4)], I.Smail[(5)], R.J.Ivison[(6)], F.Bertoldi[(7)], A.Blain[(3)], P.Cox[(2,8)], T.Greve[(3)], A.Omont[(9)]

(1) Max-Planck Institut für extraterrestrische Physik, (MPE), Garching, FRG
(linda@mpe.mpg.de, genzel@mpe.mpg.de)
(2) Institut de Radio Astronomie Millimetrique (IRAM), St.Martin d'Heres, France
(neri@iram.fr, Pierre.Cox@iram.fr)
(3) Astronomy 105-24, California Institute of Technology, Pasadena, CA 91125 USA
(awb@astro.caltech.edu, schapman@irastro.caltech.edu, tgreve@submm.caltech.edu)
(4) Department of Physics, University of California, Berkeley, CA, 94720 USA
(5) Institute for Computational Cosmology, Durham University, Durham, UK
(Ian.Smail@durham.ac.uk)
(6) Astronomy Technology Centre, Royal Observatory, Edinburgh, UK
(rji@roe.ac.uk)
(7) Radiastronomisches Institut der Universität Bonn, Bonn, FRG
(bertoldi@astro.uni-bonn.de)
(8) Institut d'Astrophysique Spatiale, Université de Paris Sud, Orsay, France
(9) Institut d'Astrophysique der Paris, CNRS & Université Pierre et Marie Curie, Paris, France
(omont@iap.fr)

## ABSTRACT

We present ~0.6" resolution IRAM PdBI interferometry of eight submillimeter galaxies at z~2 to 3.4, where we detect continuum at 1mm and/or CO lines at 3 and 1 mm. The CO 3-2/4-3 line profiles in five of the sources are double-peaked, indicative of orbital motion either in a single rotating disk or of a merger of two galaxies. The millimeter line and continuum emission is compact; we marginally resolve the sources or obtain tight upper limits to their intrinsic sizes in all cases. The median FWHM diameter for these sources and the previously resolved sources, SMMJ023952-0136 and SMMJ140104+0252 is ≤0.5" (4 kpc). The compactness of the sources does not support a scenario where the far-IR/submm emission comes from a cold ($T<30$ K), very extended dust distribution. These measurements clearly show that the submillimeter galaxies we have observed resemble scaled-up and more gas rich versions of the local Universe, ultra-luminous galaxy (ULIRG) population. Their central densities and potential well depths are much greater than in other z~2-3 galaxy samples studied so far. They are comparable to those of elliptical galaxies or massive bulges. The SMG properties fulfill the criteria of ‘maximal’ starbursts, in which most of the available initial gas reservoir of $10^{10}$-$10^{11}$ $M_{\odot}$ is converted to stars on a time scale ~3-10 $t_{dyn}$~a few $10^7$ years.

[1] Based on observations obtained at the IRAM Plateau de Bure Interferometer (PdBI). IRAM is funded by the Centre National de la Recherche Scientifique (France), the Max-Planck Gesellschaft (Germany), and the Instituto Geografico Nacional (Spain).

# 1. Introduction

Studies of the extragalactic far-IR/submillimeter background have shown that about half of the cosmic energy density comes from distant, dusty starbursts and AGN (Puget et al. 1996; Pei, Fall, & Hauser 1999). Mid-IR (ISO, Spitzer) and submillimeter (SCUBA, MAMBO) surveys suggest that this background is dominated by $z>1$ luminous and ultra-luminous infrared galaxies (LIRGs/ULIRGs: $L_{IR}$~$10^{11-13}$ $L_{\odot}$, e.g. Smail, Ivison & Blain 1997; Hughes et al. 1998; Eales et al. 2000; Bertoldi et al.2002; Elbaz et al.2002; Cowie, Barger & Kneib 2002; Perez-Gonzalez et al. 2005). During the last few years significant progress has been made in elucidating the physical properties of these submillimeter galaxies (SMGs) (e.g. Blain et al. 2002). Recently Chapman et al. (2003a, 2005) have been able to obtain rest-frame UV redshifts for 73 SMGs detected with the SCUBA and MAMBO cameras at 850µm and 1.2mm, and Swinbank et al. (2004) have subsequently obtained rest frame optical redshifts for ~20 of these. This effort has been aided to a large extent by more precise positions derived from deep 1.4 GHz VLA observations of the same fields (Ivison et al. 2002; Chapman et al. 2003a). Starting from these results we have been carrying out a survey at the IRAM Plateau de Bure interferometer, where we successfully detected molecular emission and determined line profiles in the CO 3-2 and 4-3 rotational lines for eight of these SMGs (Neri et al. 2003, Greve et al. 2005). Including data presented here, there are now 14 SMGs with published CO detections (e.g. Frayer et al. 1998, 1999, Andreani et al. 2000; Downes & Solomon 2003, Genzel et al. 2003; Greve et al. 2005; Sheth et al. 2004; Kneib et al. 2005). Here, we report for the first time sub-arcsecond resolution millimeter imaging of six SMGs, including 2 new CO detections, again undertaken with the IRAM Plateau de Bure interferometer. The main goal of this program is to determine source sizes, dynamical masses and mass densities of these sources: key parameters for determining the nature of the submm population.

## 2. Observations

The observations were carried out in winter 2003/2004 and 2004/2005 with the IRAM Plateau de Bure interferometer, which consists of six 15m-diameter telescopes (Guilloteau et al. 1992). We observed four sources previously reported in Neri et al. (2003) and Greve et al. (2005), and in addition, two new sources detected since in our survey in the extended B and A configurations under excellent weather conditions. The observations were made with baselines ranging from 64 to 408 meters. We then combined these new data with our previous B, C and D configuration observations, where the shortest baselines were 24 meters (Neri et al. 2003, Greve et al. 2005), to maximize the sensitivity and UV coverage of the maps, and to ensure that no flux was resolved out at size scales up to 6-8”. In the following, the total integration times include the lower resolution C/D configuration observations. The correlator was configured for line and continuum observations to cover simultaneously 580 MHz in each of the 3mm and 1.3mm bands. For SMMJ044307+0210 (*N4, SMMJ04431+0210, z=2.509)* the total on-source integration time was 33 hours (19 hours in A and B configurations); for SMMJ094303+4700 (H6, H7, z=3.35, Cowie, Barger & Kneib 2002, Ledlow et al. 2002, Neri et al. 2003) the total on-source integration time was 34 hours (21 hours in A and B configurations). For SMM J163650+4057 (N2 850.4, z=2.39, Ivison et al. 2002, Smail et al. 2003, Neri et al. 2003, Chapman et al. 2005) the total on source integration was ~24 hours (12 hours in A) and for SMM J163658+4105 (N2 850.2, z=2.45, Ivison et al. 2002, Greve et al. 2005, Chapman et al. 2005) the total integration time was 31 hours (16 hours in A and B). For the sources in the Hubble Deep Field North, SMM J123549+6215 (HDF 76, z=2.20, Swinbank et al. 2004; Chapman et al. 2005) and SMM J123707+6214 (HDF 242, z=2.49, Borys et al. 2004, Swinbank et al. 2004; Chapman et al. 2005) the total integration times were 31 hours (20 hours in A and B) and 38 hours (28 hours in A and B), respectively. We calibrated the data using the CLIC program in the IRAM GILDAS package (Guilloteau & Lucas 2000). Passband calibration used one or more bright quasars. Phase and amplitude variations within each track were calibrated out by interleaving reference observations of nearby quasars every 20 minutes, resulting in absolute positional accuracies of better than ±0.2 arcsec. The overall flux scale for each epoch was set on MWC 349 (1.05 Jy at 102 GHz (2.9 mm) and 1.74 Jy at 238 GHz (1.3 mm)). After flagging bad and high phase noise data, we created data cubes in natural and uniform weighting with the GILDAS package. The resulting spatial resolutions were ~1.5” and 0.6” for the 3mm and 1.3 mm bands, respectively, depending on UV coverage and source declination.

# 3. Results

Source integrated spectra, 3mm/1.3mm line and/or continuum maps of the six SCUBA sources are shown in Figs.1 through 7. Positions and derived source properties are listed in Table 1[2]. In all sources these high resolution observations let us determine source sizes (FWHM), or upper limits to them. Toward two of the SCUBA sources we find clearly detected multiple source structure. In SMMJ094303+4700 (Figs. 1 and 3) there are two millimeter sources at z~3.35, separated by 4” (~28 kpc) and likely part of an interacting group (Ledlow et al. 2002; Neri et al. 2003). The CO 3-2 emission in SMMJ123707+6214 (Figs.1 and 5) comes from a double source with a separation of ~2.5”. The PdBI observations of these six sources represent the first (sub)arcsecond resolution millimeter images (~0.6” FWHM) of representatives of the S(850μm)>5 mJy, radio detected submm source population. Sources at or above this flux density limit contribute about 22% of the 850μm extragalactic background (Smail et al. 2002), and at least 65±5% of those sources are radio ‘bright’ ($S_{1.4\,GHz} \geq 30$ μJy, Ivison et al. 2002, Chapman et al. 2003a,b).

## 3.1 Comments on individual sources

*SMMJ044307+0210* (N4, SMMJ04431+0210, z=2.509, Frayer et al. 2003)*:* SMMJ044303+0210 is one of the fifteen SMGs in the SCUBA Lens Survey (S(850μm)=7.2mJy, Smail, Ivison & Blain 1997; Smail et al. 2002). It is located behind the z=0.18 cluster MS0440+02. Smail et al. (1999) identified the submm

[2] Throughout this paper we adopt a flat ΛCDM cosmology ($\Omega_\Lambda$=0.7) with Hubble constant 70 km/s/Mpc (h=0.7)

source with the K=19.4 extremely red object (ERO) N4 about 3" north-west of an edge-on cluster spiral galaxy N1 (N4: R-K=6.3). Neri et al. (2003) detected CO 3-2 emission from N4 centered at z=2.5094. The rest frame optical line ratios suggest that N4 is a composite starburst/narrow line AGN (Frayer et al. 2003). Smail et al.(1999) estimated the foreground lens magnification from a lens model of MS0440+02 and including the impact of the nearby spiral galaxy N1 to be 4.4, approximately in the north-south direction. The uncertainties in the lensing model are ~20%, on the order of the calibration uncertainties of both the SCUBA (Smail et al. 1999) and our PdBI measurements. With this magnification, the intrinsic 850 μm flux density of the source is 1.6 mJy and the intrinsic far-IR luminosity is ~$3 \times 10^{12}$ $L_{\odot}$. This luminosity is lower than in the other sources of our sample. It is near the peak of the SMG luminosity function, where most of the far-IR luminosity emerges (Chapman et al. 2005, Smail et al. 2002), and more typical of luminosities found in local ULIRGs.

The CO 3-2 profile in Figure 1 is double-peaked with a FWHM linewidth of ~350 km/s. This emission peaks ~0.4"±0.5" south-west of the optical peak of N4 (Fig.2, Neri et al. 2003). In Figure 2 we show the CO contours superposed on a recent image of the integrated Hα/[NII] + continuum emission, obtained with the SINFONI integral field spectrometer on the ESO VLT (Förster Schreiber et al. 2005, in prep.). We determine an upper limit to the FWHM CO 3-2 size of 0.8", corresponding to a source plane limit of <0.18" (north-south) x <0.8" (east-west), or <1.5 x 6.4 kpc. The blue- and redshifted peaks of the line emission are separated by 0.7"±0.3" in the north-south direction (0.15"±0.07" in source plane). CO 7-6 emission is detected with a flux density ratio of 7-6 to 3-2 line emission of ~0.7 but given the near-equatorial location of the source our data are not good enough to construct a good quality map. We obtain a 3σ upper limit to the image plane 1.3mm continuum flux of 1.2 mJy, consistent with extrapolating from the SCUBA measurement with reasonable values of $\beta$~1.5-2.

*SMMJ094303+4700 (z=3.346):* SMMJ094303+4700 was first identified by Cowie, Barger & Kneib (2002) in a deep SCUBA map of the z=0.41 cluster Abell 851 ($S_{850}$=10.5mJy). Ledlow et al. (2002) found two radio sources H6 (S(1.4GHz)=72μJy) and H7 (55 μJy), which they proposed were associated with the SCUBA source. H6 appears to be a UV bright, narrow line Seyfert 1 galaxy at a redshift of z=3.350 from [OIII] (Takata et al. 2005, in prep) and 3.349 from Lyα (Ledlow et al. 2002), and H7 has a redshift of 3.347, also from [OIII] (Takata et al. 2005, in prep.). Cowie et al. (2002) estimated the foreground lens magnification to be 1.2. Neri et al. (2003) observed the source with the PdBI at a resolution of 3.6"x6.6" and detected strong CO 4-3 emission at z=3.346 from the weaker radio source H7, but not from H6. The new 3mm data (Figs.1 and 3, 1.9"x1.6" FWHM at PA=79°) confirm and substantially improve the CO 4-3 detection on H7. The line profile has a FWHM of 420 km/s and exhibits a double-peaked profile centered on a redshift of 3.346, the same as that found from previous CO observations (Neri et al. 2003). We tuned the 1mm receivers to the redshifted frequency of the CO 9-8 line, but did not detect this emission at a 1σ limit of 0.3 mJy averaged over the band. In our 1.3mm continuum data (Fig.3 right inset, 0.74"x0.63" FWHM at PA=64° resolution) we detect both H7 and H6, at comparable flux densities (Table 1), and both sources are unresolved. We place a (2σ) upper limit to their sizes of about 0.5" (FWHM, 3.1 kpc). The intrinsic CO 4-3 source size is 0.6"±0.2" FWHM (3.7±1.2 kpc). In our CO 4-3 channel maps we also appear to detect redshifted emission to occur a few arcsec to the north-east from H7, toward H6. This redshifted line emission is also visible as a separate weak hump in the source averaged spectrum (Fig.1). Although this detection is marginal (~3σ) its velocity does coincide approximately with that of the Lyα/CIV/NV/[OIII] line emission from H6. The velocity and possible spatial extent of this red component may thus strengthen the case for a physical connection between the two sources. The reason for the lack of CO 4-3 emission from H6, therefore, is likely that it is intrinsically CO poor, and the dust could be mainly heated by the AGN  The [OIII] emission is spatially extended, and thus the z=3.350 redshift derived from this emission is likely systemic (Takata et al. 2005, in prep).  The CO 4-3 redshift of H7 is 3.346, indicating a velocity difference of ~200 km/s (H6-H7) between the 2 sources.

*SMMJ123549+6215* (also known as HDF 76, z=2.2021, Chapman et al. 2005). Swinbank et al. (2004) determined a more precise systemic redshift of 2.203 from Hα observations in this 8.3mJy SCUBA source in the extended Hubble Deep Field North (Chapman et al. 2005). The rest frame R-band spectrum exhibits a broad Hα line, strong [NII] superposed on a broad base emission,  all indicative of the presence of an AGN (Swinbank et al. 2004). We detected strong and wide (FWHM 600 km/s) CO 3-2 emission centered at z=2.202 (Fig.1) and centered on the optical galaxy within the combined astrometric uncertainty (±0.6", Fig.4 left inset). The line profile (Fig.1) appears to exhibit two peaks on a broader base. CO 6-5 emission (also double peaked) is also clearly detected at an average 6-5/3-2 flux density ratio of ~1.5 (Fig.1) and has an intrinsic size of 0.3"±0.2" FWHM (2.5±1.6 kpc).  We marginally detect (3σ) the 1.3mm continuum at a flux density of 2 mJy from the lower sideband at 212.9 GHz, consistent with extrapolation from the SCUBA flux (Chapman et al. 2005) as discussed above.

*SMMJ123707+6214* (also known as HDF 242, Chapman et al. 2004, 2005): This 9.9mJy SCUBA source in the extended Hubble Deep Field North map of Borys et al. (2004) exhibits Hα emission of moderate width (350 km/s) centered at z=2.490 (Swinbank et al. 2004). Strong [NII] emission may indicate the presence of an AGN. We detect moderately broad CO 3-2 emission (430 km/s FWHM, Fig.1) from a source that appears to be double with a separation of ~2.5" (20 kpc, Fig.5). The south-western CO peak is coincident with the optical and radio peak, which shows complex small scale (~0.3") structure in HST (Swinbank et al. 2004) and weighted MERLIN+VLA radio images (Chapman et al. 2004). The optical map does not show emission from the weaker, north-eastern CO 3-2 source that has approximately the same redshift as the south-western peak, but very faint radio does trace the CO emission in the north-east direction. The VLA image also shows additional radio blobs ~4" north-east and 1.5" south of the main peak indicating an overall complicated source structure. We determine from UV fits CO 3-2 FWHM source sizes of 0.9"±0.3" (7.3±2.4 kpc) and <0.5"(2σ, 4 kpc) for the south-western and north-eastern peaks, respectively. This is clearly a second example (in addition to SMMJ0994303+4700) of an SMG in a complex environment with several potentially interacting components. No 1.3mm continuum flux is detected at the 3σ limit of 1.4 mJy. This limit is lower than the expected flux extrapolating from the 9.9 mJy 850 μm SCUBA flux (Table 1), and could be due either to phase decorrelation or to the presence of an extended dust component. Since our interferometer measurements should faithfully detect structures of~5" at 1mm, we consider the former explanation to be the more likely one in this source.

*SMM J163650+4057 (z=2.384).* This source (also known as SMMJ16368+4057 and Elais N2 850.4) was identified by Ivison et al. (2002: $S_{850}$=8.2 mJy) from the 8mJy SCUBA blank field survey of the Elais N2 field (Scott et al. 2002) with a bright (220μJy) 1.4 GHz radio source. Optical spectra from Chapman et al. (2003a) and Smail et al. (2003) show bright Lyα, NV, CIV, [OII] and [OIII] emission with a complex spatial and velocity structure. There is no evidence for gravitational lensing of this source (Swinbank et al. 2005). Neri et al. (2003) detected broad CO 3-2 emission from N2 850.4 centered at z=2.3853. Fig.1 shows our improved CO 3-2 line profile. It is characterized by a very broad and asymmetric line profile (FWHM ~710 km/s, FWZI 1400 km/s) with two unequal, broad humps on the blue and redshifted sides. Fig.1 also shows the CO 7-6 line data, which cover only the blue component. Within the coverage of our data, CO 7-6 and 3-2 lines exhibit similar shapes. The average CO 7-6/3-2 flux density ratio is about 1.3. Fig.6 displays our high quality 3-2 map at FWHM resolution 1.9"x1.5" at p.a. $26^{o}$ (left inset), as well as the 0.74x0.56" FWHM (at p.a.=26°) resolution map of the core of the bright 1.3mm CO 7-6 emission (right inset). Both maps are superposed on the R-band ACS image of Swinbank et al. (2005). An elliptical Gaussian fit to the UV data of the CO 3-2 emission gives a FWHM size of 0.9"x<0.4"±0.2" (7.3x<3.3±1.6 kpc) along p.a. $50\pm10^{o}$. The 7-6 core emission has a FWHM size of 1.0"x0.3" ±0.2" (8.1x2.4±1.6 kpc) along p.a.=54±10°. Channel maps of the CO 3-2 emission set an upper limit to the separation between the redshifted and blueshifted parts of the profile of about 0.4" (3.3 kpc), consistent with the size of the higher resolution CO 7-6 map. We lack the resolution to determine conclusively whether the extension along the major axis is actually two compact sources or a single, more extended source. The 1.3mm continuum flux density is 2.6 mJy, consistent with an extrapolation from the SCUBA measurement at 850 μm. Recently Swinbank et al. (2005) have published sensitive, optical and NIR integral-field spectroscopic observations of SMM J163650+4057 to probe the dynamics in the Lyα and Hα emission lines. From the Hα velocity field these authors find three distinct components (Fig.6 right inset), with the main component (B in the paper of Swinbank et al.) elongated roughly N-S. The Hα redshifts of their components A and B are consistent with those found here and in Neri et al. (2003) for the "blue" peak of the CO (3-2) and (7-6) emission, whereas their component C is consistent with the redshift of the "red" CO peak. Qualitatively, within the astrometric uncertainties the CO emission is coincident with the R-band ACS, H-band NICMOS and Hα emission from components B and C as presented in Swinbank et al. However, the orientation of the CO, at p.a.~54°, is somewhat different from the more N-S extension of the optical emission within the uncertainties of our elliptical Gaussian fit to the CO source.

*SMM J163658+4105 (*also known as *Elais N2 850.2, SMM J16370+4105 and SMM J16366+4105: z=2.453).* The submillimeter source SMMJ163658+4105 ($S_{850}$=10.7 mJy) was discovered as part of the SCUBA 8mJy 'blank field' survey of the Elais N2 field (Ivison et al. 2002; Scott et al. 2002). It is associated with an extremely red object (ERO: K=19.77, R-K=5.65, I-K=4.71, Ivison et al. 2002) and a 92(±16) μJy 1.4GHz radio source (RA=16 36 58.185 Dec=41 05 23.76). A spectroscopic redshift of z=2.43 was established from rest frame UV and optical spectroscopy by Chapman et al. (2005). Greve et al. (2005) detected the source in CO 3-2 with the PdBI in summer 2003 and found a broad line with a possible double peak, centered at a redshift of z=2.454. Our new observations clearly confirm that the CO 3-2 profile is double-peaked, with two maxima ±230 km/s around z=2.4530 and a FWHM width of ~800 km/s (Fig.1). Fig.7 (left inset) shows the integrated CO 3-2 map at 1.6"x1.2" FWHM resolution and the right inset shows the integrated CO 7-6 emission at 233 GHz that dominates the upper sideband of our new high resolution data set (0.7 x 0.56" FHWM at PA= -169°). No significant continuum emission is detected at the 3σ level of 1.5 mJy, which is likely due to phase decorrelation or some extended dust emission. A circular Gaussian fit to the UV data of the CO 7-6 line gives a source size of 0.4”±0.15” (FWHM, 3.2±1.2 kpc), indicating compact emission even with the limited signal-to-noise of the data set. In position-velocity diagrams of the CO 3-2 emission (~2” resolution) the positions of blue and red maxima of the double-peaked profile are identical to within ~0.4”, consistent with the compact size of the CO 7-6 emission. Within the absolute astrometric accuracy of the radio-optical reference frames (±0.3”), the compact submillimeter source, the 1.4GHz VLA source and the weak, spotty K-band emission of the ERO are probably all positionally coincident.

# 4. Discussion

## 4.1 SMGs are scaled-up versions of local ULIRGs

### *4.1.1 SMGs are compact*

Fig.8 summarizes what we presently know about the intrinsic sources sizes of the SMG population, adding to the sources observed here the information on the two lensed sources SMM023952-0136 (Genzel et al. 2003) and SMM140104+0252 (Ivison et al. 2001; Downes & Solomon 2003). For SMM140104+0252 we adopt a low amplification model of 3.5 to describe the intrinsic properties of this source (e.g. Smail et al. 2005; Baker et al. 2005, in prep). The basic result is that the mm line (and continuum emission where we have had sufficient sensitivity at 1mm to measure it) in these 10 representatives of the SMG population is compact. The distribution of the intrinsic FWHM sizes of these sources, as well as the intrinsic separation of the blue- and red-shifted CO 3-2 components in SMM J16359+6612 (Kneib et al. 2005) is shown as a hatched histogram in Figure 9. The lensing corrected, median FWHM diameter of our SMG sample is ≤0.5", with a dispersion of 0.2". This corresponds to a linear diameter of ≤4±1.6 kpc.

Chapman et al. (2004) have carried out ~0.3" imaging with MERLIN/VLA of the 1.4GHz emission of 11 spectroscopically identified SMGs in the range z=1.9-2.7. They find a median (half-light) source size of 0.4±0.1" corresponding to a linear diameter of 3.2±0.8 kpc. The distribution of these sizes is shown in Figure 9. Assuming that the radio emission does serve as a proxy of the far-infrared/submillimeter emission (because of the radio-IR relationship) these results are in excellent agreement with the FWHM size estimations from our CO and 1mm continuum data. Recent Spitzer observations show that the local radio-far-IR correlation does indeed hold to z≥1 (Appleton et al. 2004). There is even evidence for the FIR-radio correlation for redshifts up to z~6 for the starburst component of radio quiet quasars (Beelen et al. 2005). Spatially resolved studies of radio and CO emission regions in local ULIRGs show that the FWHM emission sizes are also within a factor ~2 of each other. For the 5 bright local ULIRGs imaged at ~1-2" resolution in both CO and 1.49 GHz radio continuum (Downes and Solomon 1998, Condon et al. 1991) the mean FWHM equivalent sizes are 1.3±0.5" (~700 pc) and 1.0±0.4" (~500 pc), respectively. For these local galaxies both the molecular and radio continuum emission are largely concentrated within the central kpc.

#### *4.1.2 Molecular gas in SMGs is dense*

In concluding that the CO emission is compact we need to take into account that in three of the ten SMGs the tracer used to estimate the size was the excited CO 6-5 or 7-6 rotation line (requiring warm (≥35K) and fairly dense ($\geq 10^3$ $cm^{-3}$) gas). How representative is this excited gas for the molecular interstellar medium as a whole? The 6-5/3-2 line flux density ratio in SMM123549+6215 is ~1.5, the 7-6/3-2 ratio in SMM044307+0210, SMM 140104+0252 (Downes & Solomon 2003), SMM16350+4057 and SMM 163658+4105 ranges between 0.7 and 2 and the 9-8/4-3 ratio in SMM094303+4700 is <0.2. We carried out large velocity gradient, molecular excitation models to explore the physical parameters of the molecular gas implied by these line ratios. We adopted a $CO/H_2$ fractional abundance of $\sim 10^{-4}$, a local line width of ~50-100 km/s, similar to conditions in local ULIRGs and our own Galactic Center, and a gas temperature ~35-50K, consistent with Blain et al. (2004) and our discussion in the next section. We find that local molecular hydrogen densities of about $10^4$ $cm^{-3}$ are required to account for the above line ratios, fairly insensitive to gas temperature. While significantly higher than the volume averaged densities derived from molecular gas mass and source diameter (section 4.2 and Table 2), these densities are comparable to the local densities derived from HCN/CO line ratios in ULIRGs and LIRGs (Gao & Solomon 2004). The CO 6-5 or 7-6 line shapes in SMMJ123549+6215 and SMM163650+4057 are similar to those of the lower J-lines (Fig.1), a trend which has recently been found also in other high-z galaxies by Weiss et al. (2005, in prep.). In SMMJ140104+0252 Downes & Solomon (2003) find similar sources sizes for CO 3-2, CO 7-6 and 1.3mm continuum. We conclude that the CO 7-6 or 6-5 line emission can be plausibly used as a tracer of the global CO emission in the SMGs. In the Milky Way there is a strong correlation between sites of active star formation and concentrations of dense ($\geq 10^4$ $cm^{-3}$) molecular gas (Lada & Lada 2003). The fairly high global gas densities implied by the strong 6-5 or 7-6 CO lines in SMGs thus would suggest that star formation must be efficiently proceeding throughout these systems. We will discuss this aspect in the next section.

### *4.1.3 Cold dust models are not supported*

We now proceed with the assumption that the average characteristic radius of powerful SMGs (at half intensity) is $<R_{1/2}>$~2 kpc. Naturally there likely is gas present on scales greater than $R_{1/2}$, and this is especially true for those sources with complex structure, such as *SMM J163650+4057*. However, the quality of our present mm-interferometry is not good enough to derive 'isophotal' or 'half-mass' radii that would likely be larger than $R_{1/2}$. We believe that the average densities, column densities and temperatures derived below are well described by using $R_{1/2}$ as a characteristic size. Dynamical masses for the baryonic component estimated with $R_{1/2}$ are lower limits.

For a spherically symmetric dust emission source absorbing and re-radiating as a modified grey-body (temperature $T_d$, frequency dependent emissivity $Q(\nu)\sim\nu^{\beta}$) all of the intrinsic short-wavelength radiation of luminosity $L_{bol}$, the radius R of the dust source depends on luminosity in the usual modified Stefan-Boltzmann law

$$R = \left( \frac{L_{bol}}{4\pi f(T_d, \beta)} \right)^{1/2} \qquad (1)$$

where $f\sim T_d^{4+\beta}$. In this simple model high-z SMGs are similar to local Universe ULIRGs (in the sense that their characteristic dust temperature and dust properties are similar) if $R/(L_{bol})^{1/2}$ are approximately the same in both types of sources. Local ULIRGs have median CO (1-0/2-1) radii of 0.6 (±0.3) kpc (Downes & Solomon 1998, Bryant & Scoville 1999) and median luminosities of $10^{12.1\pm0.1}$ $L_{\odot}$, resulting in $R/L^{1/2}$= $10^{-6.3\pm0.3}$ kpc/$(L_{\odot})^{1/2}$. The SMGs in this paper and in the samples of Neri et al. (2003) and Greve et al. (2005) have a median luminosity of $10^{13.1\pm0.14}$ $L_{\odot}$, resulting in $R/L^{1/2}$=

$10^{-6.4\pm0.25}$ kpc. The luminosity surface densities of local ULIRGs and our observed SMGs are identical: SMGs are scaled-up ULIRGs.

Kaviani, Haehnelt & Kauffmann (2003) and Efstathiou & Rowan-Robinson (2003) have proposed that high-z SMGs are much colder ($T_d$~20-30 K) and thus more spatially extended (R~5-20 kpc) than local ULIRGs ($T_d$~40 K). SMGs could then be akin to local, lower luminosity galaxies with far-IR emission dominated by cool extended 'cirrus' dust, with about 4 times lower luminosities than in the ULIRG model, for a fixed 850μm/1.3mm flux density. In this model, source radius R, the flux density at observed frequency ν, $S_\nu$ and the bolometric source intensity $I_{bol}=L_{bol}/(4\pi R^2)$) are related through (Kaviani et al. 2003)

$$R\, S_\nu^{-1/2} = (1+z)^{3/2}\, D_A\, (g_{\nu(1+z)} I_{bol})^{-1/2} \quad (2),$$

where $D_A$ is the angular diameter distance (~1.7 Gpc at z~2-3) and $g_{\nu(1+z)}$ is a normalization function relating the bolometric luminosity to the luminosity density at ν(1+z) ($g_{\nu(1+z)}=L_{\nu(1+z)}/L_{bol}$). Kaviani et al. compute R $S_{350GHz}^{-1/2}$ for both the cold cirrus and for the warm ULIRG models. In the cirrus models source radii for sources of $S_{850}$= 5-10 mJy at z~2-3 should be about 2" - 3" (~15-20 kpc), a factor 10 greater than what we observe in the sources studied here. In most of the sources, our 1.3mm continuum detections are consistent with extrapolations from the SCUBA 850 μm measurements for β values of 1.5 or 2, indicating that we are not missing "extended flux".

Blain et al. (2004) have computed dust temperatures for 73 SMGs with spectroscopic redshifts from the Chapman et al. (2005) sample, using their 850μm flux densities and the local Universe, radio-far-IR relation (to calculate far-IR luminosities). The derived temperatures for $L\sim10^{13}\,L_\odot$ (and for β=1.5) are 39±3 K, again in excellent agreement with the scaled-up ULIRG model. Further support for this model comes from the ratio of the 850μm to 1.4 GHz flux density ratios of SMGs, and by recent 350 μm photometry of SMGs by Kovacs et al (2005, in prep.). The observed and derived properties of the SMGs listed in Table 2 are consistent with the scaled-up ULIRG model, including in addition, a larger gas fraction than the z~0 systems.

#### *4.1.4 Gas motions are ordered*

The CO 3-2/4-3 line profiles in 5 of the sources we have studied here show a double peaked profile characteristic of orbital motion, either of a single rotating disk or of two galaxies in an ongoing merger system (Fig. 1). Overall of the 14 SMGs with CO profiles 8 (57%) have double peaked profiles (Neri et al. 2003, Greve et al. 2005). For comparison, at least 33% of 36 local ULIRGs studied in CO 1-0 emission by Solomon et al. (1997), have double-peaked or flat-topped spectra, consistent with the SMGs given the limited S/N and small number statistics of both samples.

We have modeled the line profiles and sizes of the 3 galaxies with the highest S/N CO spectra, SMMJ023952-0136, SMMJ094303+4700 and SMMJ163658+4105 (Fig. 1 and Genzel et al. 2003), with simple disk models convolved with our spectral and spatial resolution (see also description of disk model in Genzel et al. 2003). In the framework of this model, the shapes of the line profiles are sensitive to the ratio of local (random) velocity dispersion ($\sigma_g$) to global circular/rotation velocity ($v_c$). To fit the line profiles in these three galaxies we find $\sigma_g/v_c$ from 0.15-0.4.

While a merger model appears most plausible, in analogy to local ULIRGs, the data are not yet of sufficient angular resolution to confirm (or refute) merger or simple disk scenarios. Both models have similar physical predictions for cold gas that can dissipate on a dynamical timescale. From rest-frame UV morphologies, colors and spectra there is substantial evidence that SMGs live in a complex environment (Smail et al. 2002, 2003, Chapman et al. 2003c, Greve et al. 2003). Obviously the two sources H6 and H7 in SMMJ094303+4700 and the two components of SMMJ123707+6214 (HDF 242) appear to be part of interacting systems on scales of ~25 kpc. Likewise, the components J1 and J2 in SMM140104+0252 are physically associated on a scale of ~20 kpc (Ivison et al. 2001, Tecza et al. 2004). The high resolution mm-data from the SMGs we have observed thus far, taken together with published SMMJ023952-0136 and 140104+0252 data (Genzel et al 2003; Downes & Solomon 2003), show that there is a mix between compact sources and more widely spaced, perhaps intermediate stage mergers with nuclear separations ≥5 kpc. Over 80% of local ULIRGs with log L > $10^{12.5}$ are compact, single nucleus systems or advanced mergers with nuclear separations ≤2 kpc, with the remainder being at earlier merging stages, (Veilleux, Kim & Sanders 2002). The morphological properties of the SMGs are again qualitatively consistent with those of local ULIRGs in this regard (Chapman et al. 2003c; Pope et al. 2005). The mm maps are not of sufficient quality to discriminate between intermediate-to-advanced stage mergers or rotating disks in a single nucleus system. To investigate this point further, we have taken the observed CO 2-1 PdBI maps of Arp 220, Mrk 273 and NGC 6240 (Downes & Solomon 1998, Tacconi et al.1999), moved them to the average redshift of the SMG sample and convolved the resulting data with a 0.6" beam. In these artificial maps we are (marginally) able to detect a broadening of the beam but we cannot determine whether the sources are intrinsically double, or single. Weiss et al. (2005, private communication) have observed a number of rotational CO lines in the strongly lensed SMG SMMJ16359+6612 that shows two prominent, almost equal intensity line emission peaks (see also Kneib et al. 2005, Sheth et al. 2004). They find that the line profiles do not change between rotational quantum number 1 to 6, suggesting that the physical conditions and excitation of the molecular gas in the two peaks is virtually identical, strongly arguing for an origin in the same dynamical system and supporting a rotating disk interpretation for this SMG. We caution, however, that this source is 5-10 times fainter than those sources included in our study.

In the framework of a disk-like distribution of gas with high angular momentum, as is seen in local ULIRG mergers (Downes & Solomon 1998), the median circular velocity of the SMG sample is $<v_c>\sim400$ km/s. Using simulated gas disks as a guide, we have divided here the observed FWHM line widths by 2.4 to get $v_c \sin(i)$ and then corrected the sample median statistically for inclination by multiplying by $\pi/2$ (Neri et al. 2003, Greve et al.2005). The median dynamical mass encircled within $R_{1/2}$ then is $M_{t\,½}\sim7\times10^{10}\ M_\odot$. Since gas is very likely present outside $R_{1/2}$, this is obviously a strict lower limit to the total baryonic (stars plus gas) mass of the SMG population (Greve et al. 2005; Tecza et al. 2004). The virial theorem shows that for a merger scenario the dynamical masses are approximately a factor of two larger (Genzel et al. 2003). The median gas mass, as estimated from the CO flux I(CO) using the $N(H_2)/I(CO)$ ratio determined for local ULIRGs by Downes and Solomon (1998), including a correction for 10 % interstellar helium, and assuming $M(HI)/M(H_2)<1$ in the CO emission region is $2.4\times10^{10}\ M_\odot$. The median gas fraction $f_g$ thus is ~40%. For comparison local ULIRGs have $<R_{1/2}>=0.6$ kpc, $<M_{t\,½}>\sim6\times10^9\ M_\odot$ and $f_g\sim0.16$ (Downes & Solomon 1998). SMGs thus appear to be ~2.5 times more gas rich than local ULIRGs if the z~0 CO to $H_2$ conversion factor also applies at z>2. The ratio of the total dynamical mass within the entire region where CO is observed (several kpc) to $<M_{t\,½}>$ is about 5. Higher spatial resolution and higher sensitivity CO interferometry, such as will be possible in the future with ALMA, will be required to test whether such a large correction is also appropriate for SMGs.

## 4.2 SMGs have matter densities and well depths similar to ellipticals/ large bulges

From basic observed and derived quantities, $v_c$, $R_{1/2}$, $M_{gas}$, $\sigma_g$, and $f_g$ we can now derive the mean matter and gas volume and surface densities using

$$<\rho_m>=1.3x10^2\left(\frac{v_c}{400km/s}\right)^2\left(\frac{R_{1/2}}{2kpc}\right)^{-2}\left(\frac{\sigma_g/v_c}{0.25}\right)^{-1} \quad (cm^{-3})$$

$$<\Sigma_m>=5.9x10^3\left(\frac{v_c}{400km/s}\right)^2\left(\frac{R_{1/2}}{2kpc}\right)^{-1} \quad (M_\odot\ pc^{-2})$$

$$=5.3x10^{23}\left(\frac{v_c}{400km/s}\right)^2\left(\frac{R_{1/2}}{2kpc}\right)^{-1} \quad (cm^{-2}) \qquad (3).$$

Here we have assumed that the FWHM z-height of the star forming/molecular cloud layer, 2h/R, is two times the ratio of local velocity dispersion $\sigma_g$ to circular velocity. From the average circular velocities, radii and velocity dispersions listed in Table 2 we find that SMGs have mean volume and surface densities (Table 2) of ~100 $cm^{-3}$ and 5000 $M_\odot\ pc^{-2}$, averaged over the emission region of radius $R_{1/2}$. Gas volume and surface densities are lower by the gas fraction, $f_g$~0.4. Local ULIRGs have mean circular velocities of 200 km/s (CO: Downes & Solomon 1998, Solomon et al. 1997) to 260 km/s (stars: Genzel et al. 2001, Tacconi et al. 2002), indicating that their gravitational potential wells are somewhat smaller than SMGs. Within $<R_{1/2}>$=600 pc this corresponds to average volume and surface mass densities of 350 $cm^{-3}$ and 4900 $M_\odot pc^{-2}$, similar to SMGs. Ellipticals and Sa bulges have circular velocities, and thus potential well depths, similar to SMGs ($v_c$ ~210 to 450 km/s) and also have comparable average mass densities of ~100-200 $cm^{-3}$ (Kormendy 1989). These volume densities in turn resemble the average hydrogen densities inside actively star forming Giant Molecular Clouds (GMCs) in our Milky Way ($10^6$ $M_\odot$ within ~10-40 pc: average densities 100-300 H $cm^{-3}$; Young & Scoville 1991). Both ULIRGs and SMGs appear to contain a high filling factor interstellar medium of dense molecular gas.

## 4.3 SMGs are 'maximum' starbursts

We show in this section that the observed properties of SMGs (and local ULIRGs) can be fully and self-consistently understood in the framework of 'maximum' starbursts (Elmegreen 1999), in which a significant fraction of the available gas content is converted into stars in a few times the dynamical time scale. Following Kennicutt (1998) and Elmegreen (1999) we write this maximum ('Schmidt-law') star formation rate as

$$SFR_{\max} = \frac{\varepsilon f_g M_t}{t_{dyn}} = 630 \varepsilon_{0.1} f_{0.4} \mathrm{v}_{400}^3 \qquad [M_\odot yr^{-1}], \text{ and}$$

$$\Sigma_{\mathrm{SFR\ max}} = \frac{SFR_{\max}}{\pi R^2} = 43 \varepsilon_{0.1} f_{0.4} \mathrm{v}_{400} \Sigma_{t,5000} R_2^{-1} \qquad [M_\odot yr^{-1} kpc^{-2}] \qquad (4),$$

where $M_t$ is the total (dynamical) mass within radius R ($R_2$=R/(2 kpc)), including gas, stars and dark matter, $f_g$ is the gas fraction ($f_{0.4}=f_g/0.4$), and $t_{dyn}=R/v_c$ is the dynamical time scale. The parameter ε describes the local efficiency of star formation. Star formation in our own Galaxy can be described by an efficiency of 10% ($\varepsilon_{0.1}$=1, Elmegreen 1999). $\Sigma_{t,5000}$ is the total (gas+stars+DM) matter density within R, in units of 5000 $M_\odot pc^{-2}$. Assuming a Salpeter-like initial mass function from 1-100 $M_\odot$ (similar to a Kroupa (2001) IMF) SMGs have star formation rates of ~900 $M_\odot$ $yr^{-1}$ and star formation rate surface densities of ~80 $M_\odot$ $yr^{-1}$ $kpc^{-2}$ (Table 2). These values are similar to and even somewhat larger than these theoretical maximum star formation rates, perhaps indicating that in their very dense, high pressure ISM the star formation efficiency may be somewhat larger than in the quiescent environment of the Galaxy, $\varepsilon_{SMG}$~0.15-0.2.

Under which physical conditions would one theoretically expect such 'maximum' starbursts to occur and do SMGs (and ULIRGs) fulfill these criteria? The first criterion is that the interstellar medium becomes gravitationally unstable on a global scale in the presence of galactic differential rotation and shear (in rotating disks, Toomre 1964), or in the presence of local turbulent motions (for a spherical configuration, Spitzer 1942). Elmegreen (1999) has shown that in both disk-like and spherical configurations, large scale gravitational collapse must set in whenever the gas density becomes a significant fraction of the 'virial' density

$$\rho_{gas}(\leq R_g) \approx \rho_{vir} = \frac{M_t}{4\pi/3 \, R_g^3} \qquad (5).$$

Expressing this criterion in terms of the observable quantities we find that the threshold for large scale collapse is given by

$$\Sigma_{gas,crit} \geq 2.3x10^3 \frac{\sigma_{100}\mathrm{V}_{400}}{R_2} \quad [\mathrm{M}_\odot pc^{-2}], \text{ or alternatively}$$

$$\left(\frac{\sigma}{\mathrm{v_c}}\right) \leq 0.35 f_{0.4} \qquad (6).$$

Both of these conditions are met by the SMGs (Table 2). From these considerations and the observed properties the conclusion is inevitable that SMGs are prone to large scale star formation. Equation (4) shows that their large masses, compact sizes and high inferred gas fraction naturally lead to the very large star formation rates observed. The fact that in the framework of Toomre's disk stability parameter Q SMGs have Q~1 can be interpreted in terms of the concept of self-regulated star formation (e.g. Gammie 2001, Thompson, Quataert & Murray 2005). Gravitationally unstable disks (Q<<1) start forming stars vigorously. Star formation then increases the local sound speed/velocity dispersion σ, which in turn increases Q. Once Q is driven above 1, star formation ceases again and Q finally self-regulates at Q~1.

The second key criterion is whether or not the maximum starburst can be maintained for any length of time, given the large negative feedback from stellar winds, supernova explosions and radiation pressure acting on dust (Elmegreen 1999; Scoville 2003, Thompson et al. 2005). This (negative) feedback is now considered to be a key factor in determining the evolution of galaxies. Once this feedback sets in after a typical OB star lifetime, the pressure from supernova explosions, radiation and stellar winds will invariably terminate the starburst by dispersing and driving out the gas unless the gravitational potential well is deep enough. Based on the estimates of Meurer et al. (1997), Elmegreen (1999), Murray, Quataert & Thompson (2005) and Thompson et al. (2005) for the balance of momentum driven winds and self-gravity the maximum star formation rate that can be maintained in a spherical configuration is given by

$SFR_{\rm max,feedback} \leq 9.6x10^3\, f_{0.4} {\rm v}_{400}^4 \qquad [{\rm M}_\odot yr^{-1}]$, or equivalently

$\sigma_{\rm g} \geq 33\varepsilon_{0.15} \qquad [\text{ km/s}] \qquad (7).$

For a thin disk the situation is more complicated. Radiation pressure in the optically thick limit is more efficient in lifting the gas up from the midplane and the maximum critical star formation rates may be a few times smaller (Thompson et al. 2005) than given in equation (7). Nevertheless it would appear that SMGs (and ULIRGs) have sufficiently deep potential wells to retain their (cold) gas reservoir. If so, star formation will not be terminated by negative feedback unless, as discussed by Granato et al. (2004) and DiMatteo, Springel & Hernquist (2005), accretion onto the central black holes creates even more powerful negative feedback than star formation.

However, as shown by Alexander et al (2005), the low level of AGN activity in most SMGs may well last for the duration of the typical starburst lifetime, before the black hole has reached the mass of $\sim 10^8$ $M_\odot$ needed for powerful enough feedback. The current star formation activity in SMGs may plausibly convert much of the original gas mass of $10^{10}$-$10^{11}$ $M_\odot$ (given that we observe them typically halfway through their starburst epoch; e.g. Tecza et al. 2004) to stars within their characteristic star formation time scale, $t_{SF} \sim t_{dyn}/\varepsilon \sim 40 - 100$ Myr. This is exactly the requirement for rapidly forming ellipticals and bulges in the redshift range between 4 and 2. The condition that maximum starbursts must have a deep enough potential well to retain their gas reservoir, and therefore also their freshly made heavy elements, is in excellent agreement with the super-solar α-element abundance deduced by Tecza et al. (2004) for SMMJ140104+0252.

## 4.4 Comparison to optical tracers and UV bright z~2-3 galaxy samples

A key goal of our work is to better understand the properties and evolution of the SMG population. For this purpose, we would like to compare the kinematics, sizes and masses of the SMGs derived in this paper with those in other high-z galaxy populations. However, in the other populations there are as yet very few observations with mm tracers. Properties have been determined almost exclusively from rest-frame optical/UV lines and continuum. It is useful, therefore, to begin by comparing the kinematic/structural parameters of SMGs obtained in CO with those in optical tracers, primarily in Hα. At the present time there are six SMGs for which both mm- and optical tracers have been observed: SMM044307+0210 (Frayer et al. 2003, Förster Schreiber et al. 2005, in preparation), SMM123549+6215 (Swinbank et al. 2004), SMM123707+6214 (Swinbank et al. 2004), SMM140104+0252 (Tecza et al. 2004,

Swinbank et al. 2004), SMM163650+4057 (Swinbank et al. 2004, 2005) and SMM163658+4105 (Swinbank et al. 2004). In these six sources the ratios of CO to Hα velocity widths and sizes are 1.35(±0.6) and 0.8(±0.17) where the value in the parenthesis is the 1σ dispersion. For a sample of 15 SMGs and optically faint radio galaxies (OFRGs) in the GOODS-N field Smail et al. (2004) derive a median optical half-light radius of 0.6", somewhat larger than the median millimeter FWHM size in our sample and that of the radio data in Chapman et al. (2004). The optical line emission in SMGs may be somewhat more extended spatially and narrower in velocity width than the millimeter emission (cf. Swinbank et al. 2004). The more concentrated millimeter and submillimeter emission likely come from a denser central obscured starburst surrounded by the lower extinction Hα zone that is contributing ~10% of the star formation activity (Greve et al. 2005). This inside-out stratification is supported by the comparison of optical/UV and high resolution millimeter maps (Figure 8) and may be plausibly explained by strong extinction gradients. The UV/optical maps trace lower extinction, lower luminosity regions on the outside of a highly obscured, central starburst component. This is the reason that reliable spectroscopic redshifts from Lyα have been feasible from these very highly obscured systems (Chapman et al. 2003a, 2005).

Next we turn to a comparison between SMGs and UV bright star forming galaxies in the same redshift range. In contrast to the very dusty SMGs, UV/optical tracers in the UV bright star forming galaxies more likely delineate the bulk of the star formation in these galaxies. Erb et al. (2003, 2004) and Förster Schreiber et al. (2005, in prep.) have carried out long-slit and integral field spectroscopy of Hα in a sample of z~2-2.5 UV bright star forming galaxies (the 'BX' sample: Steidel et al. 2004). The K<20 tail of these galaxies appear to have stellar masses $\geq 10^{11}$ $M_\odot$ and solar or super-solar metallicities (Shapley et al. 2004, 2005), similar to the SMGs (Greve et al. 2005, Tecza et al. 2004, Swinbank et al. 2004). It is therefore of great interest to explore whether these galaxy samples may be related through evolution. Erb et al. (2003, 2004) and Förster Schreiber et al. (2005) derive median FWHM line widths of 260 (±150) km/s and $R_{1/2}$=0.24”-0.45” for these UV bright star forming galaxies. For the z~3 Lyman break galaxies (LBGs) Pettini et al. (2001) derive FWHM velocity widths of 160 km/s and sizes of $R_{1/2}$=0.3”. The dynamical masses of SMGs are 2-3 times larger than those of the BX population and more than an order of magnitude greater than the LBGs. Matter volume densities are an order of magnitude or more larger in the SMGs than in the z=2-3 UV population. This result does not change if only the K<20 tail is considered for which Shapley et al. (2004) derive super-solar metallicities and stellar masses comparable to those of the SMGs considered here.

# 5. Conclusions

We have carried out the first sub-arcsecond to arcsecond resolution, mm interferometry of 8 submillimeter galaxies between redshifts 2.38 and 3.35. The 1.2mm emission from dust continuum and CO line emission from the 3-2/4-3 and 6-5/7-6 lines is compact; we marginally resolve or obtain tight upper limits to the FWHM sizes in all cases. The characteristic linear source diameters are ≤4 kpc. Five of the eight sources we have studied exhibit double-peaked line profiles indicative of orbital motion with significant angular momentum, either in a rotating disk, or an advanced merger. Our data are not yet detailed enough to distinguish between these two options.

Combining our spatial information and detections of new sources with the recent results from a CO survey of submillimeter galaxies by Neri et al. (2003) and Greve et al. (2005), we find that dynamical masses, matter volume and surface densities of the SMG population, based now on 14 sources, are very large and comparable to those in massive local ellipticals and Sa bulges. We also find that the SMGs we have studied are similar to local Universe ultra-luminous infrared galaxy mergers (ULIRGs), suitably scaled for their larger masses, luminosities and star formation rates, as well as their greater gas fractions.

The SMGs we have studied can be self-consistently understood in the framework of 'maximal' starbursts, where most of the available cold gas reservoir ($10^{10}$-$10^{11}$ $M_{\odot}$) is converted to stars on a time scale 1/ε times the characteristic dynamical time scale (~40 -100 Myrs), where ε~0.15-0.2 is the local star formation efficiency of an SMG. In such a maximal starburst the gas concentration is large enough to trigger a global starburst in the presence of rotation, galactic shear and turbulence. The gravitational potential well depth is also large enough to prevent galactic outflows driven by supernovae, stellar winds and radiation pressure to disperse the interstellar material once the starburst has begun. Only the yet more powerful AGN feedback can plausibly stop the maximum starburst once triggered.

*Acknowledgements:* We thank Michael Grewing and the staff of the IRAM Observatory for their support of this program. We are also grateful to Andrew Baker, Todd Thompson, Norm Murray and Eliot Quataert for insightful discussions and comments, to Dennis Downes for helpful comments and access to his PdBI maps of local ULIRGs, and to Tom Muxlow for useful information on radio source sizes. We thank the anonymous referee for some very insightful comments, which have improved the manuscript. IRS acknowledges support from the Royal Society.

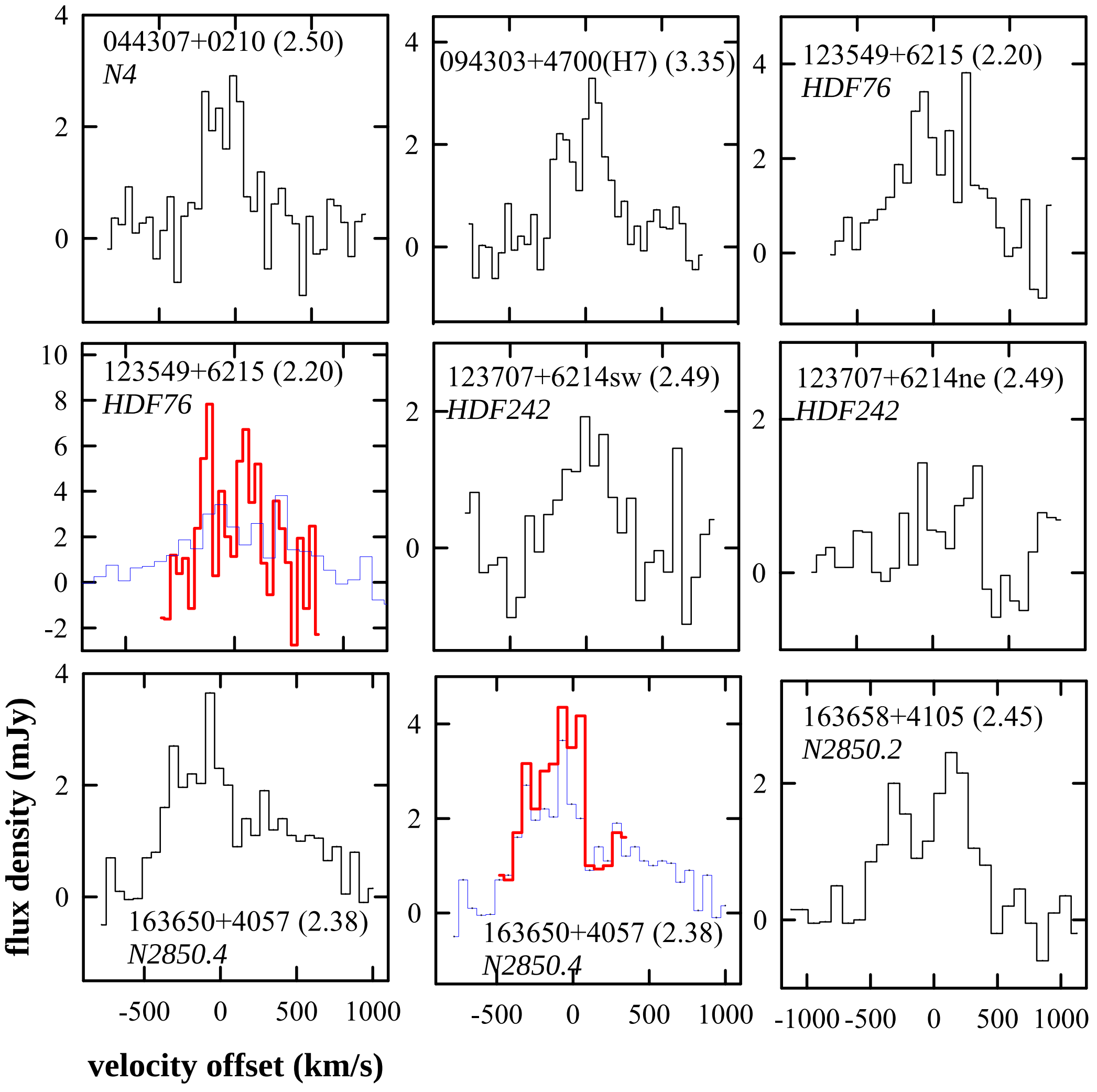

Fig.1. Integrated CO spectra of the sources. For SMMJ094303+4700 we show the CO 4-3 profile, in all other cases the CO 3-2 profile. For the left-most spectrum in the middle row and the middle spectrum in the bottom row we plot the CO 6-5 (SMMJ123549+6215) and CO 7-6 (SMMJ163650+4057) profiles (heavy lines) on top of the CO 3-2 profile (thin lines). The middle and right profiles in the middle row are the south-western and north-eastern sources of SMMJ123707+6214.

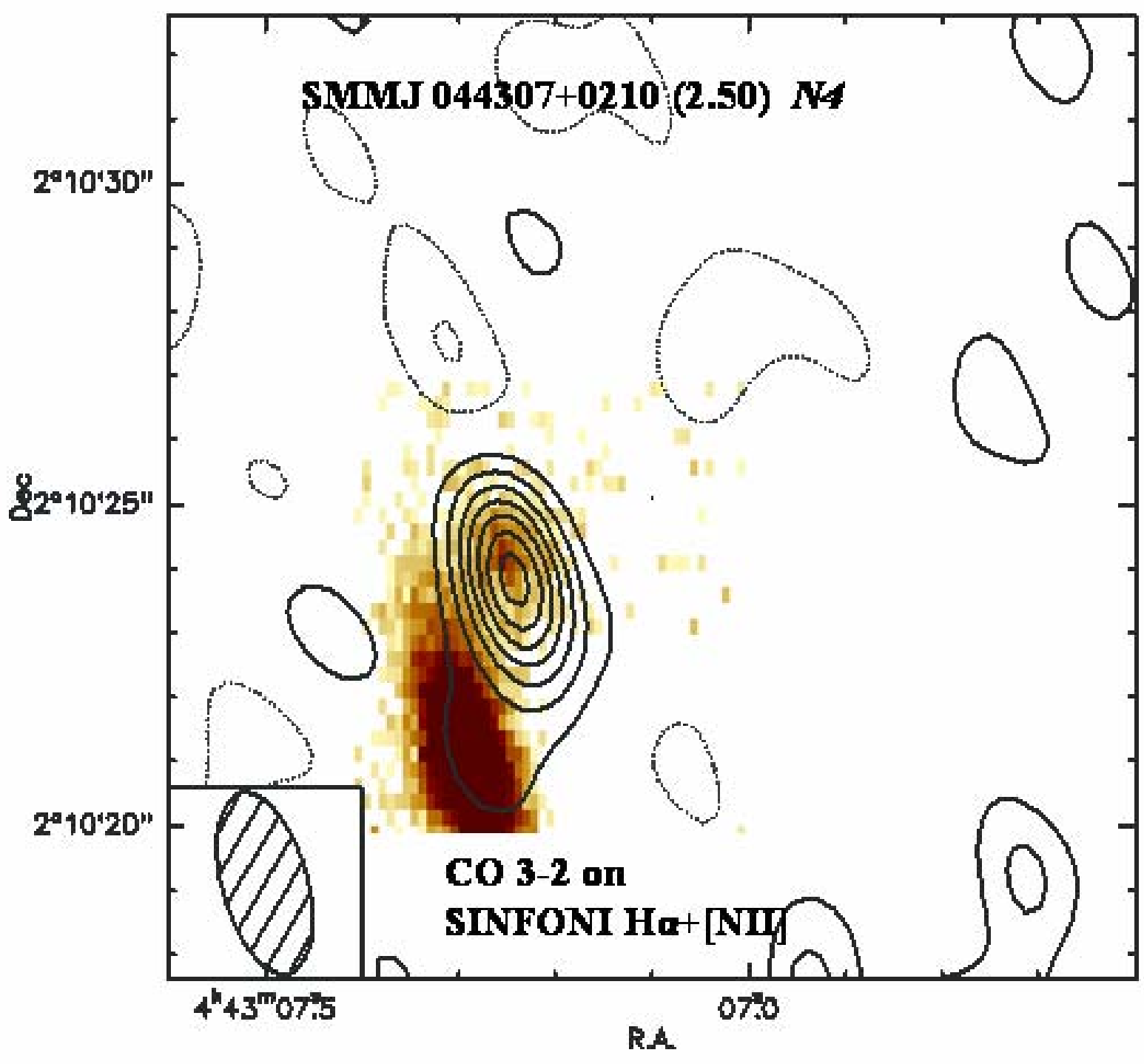

Fig.2. Integrated CO 3-2 emission in SMMJ044307+0210 (N4: z=2.50, see Frayer et al. 2003, Neri et al. 2003), superposed on the integrated Hα/[NII] + continuum emission from the source, observed with the SINFONI integral field spectrometer on the ESO VLT (grayscale, ~0.6" resolution, Förster Schreiber et al. 2005 in prep.). The edge on galaxy ~3-4" south-east of N4, N1, is in the z=0.18 foreground cluster MS0440+02. The FWHM resolution of the CO 3-2 map is 2.9"x1.3" and the contours start at 0.2 mJy/beam (1.2σ), spaced by 0.2.mJy/beam.

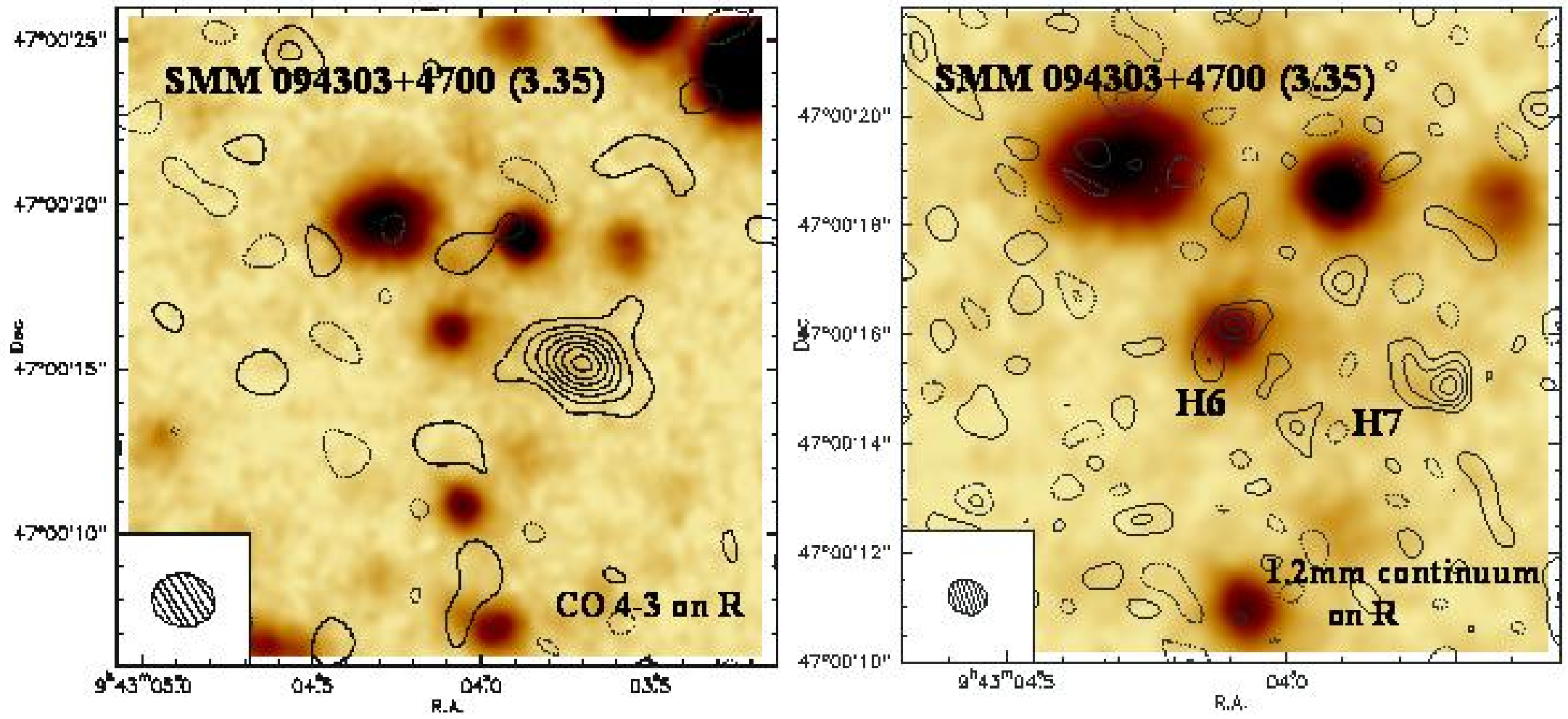

Fig.3. Integrated CO 4-3 and 1.26 mm continuum in SMMJ094303+4700 (z=3.35, see Ledlow et al. 2002, Neri et al. 2003). Left: integrated CO 4-3 emission (FWHM 1.9”x1.6” FWHM at PA=79°) superposed on an R-band image of the cluster Abell 851 (z=0.41, Kodama et al. 2001). The contours are linearly spaced, starting at 0.25 mJy/beam (~2σ). Right: Contour map of the 237.1 GHz (1.26 mm) continuum emission. The synthesized beam has a FWHM of 0.74”x0.63”: at p.a. 64$^{o}$ (lower left inset). Contours are evenly spaced starting at 0.4 mJy/beam (1.5σ). CO 9-8 at 238.6 GHz does not contribute more than about 3 mJy (1σ) to the source flux. H6 and H7 are the two radio sources in the field. The redshift of H6 from [OIII] is 3.350 (Takata et al., in prep.). The CO 4-3 redshift of H7 is 3.346, showing that H6 and H7 are located in the same physical structure with a projected separation of ~25 kpc and a velocity difference of ~200 km/s (H6-H7).

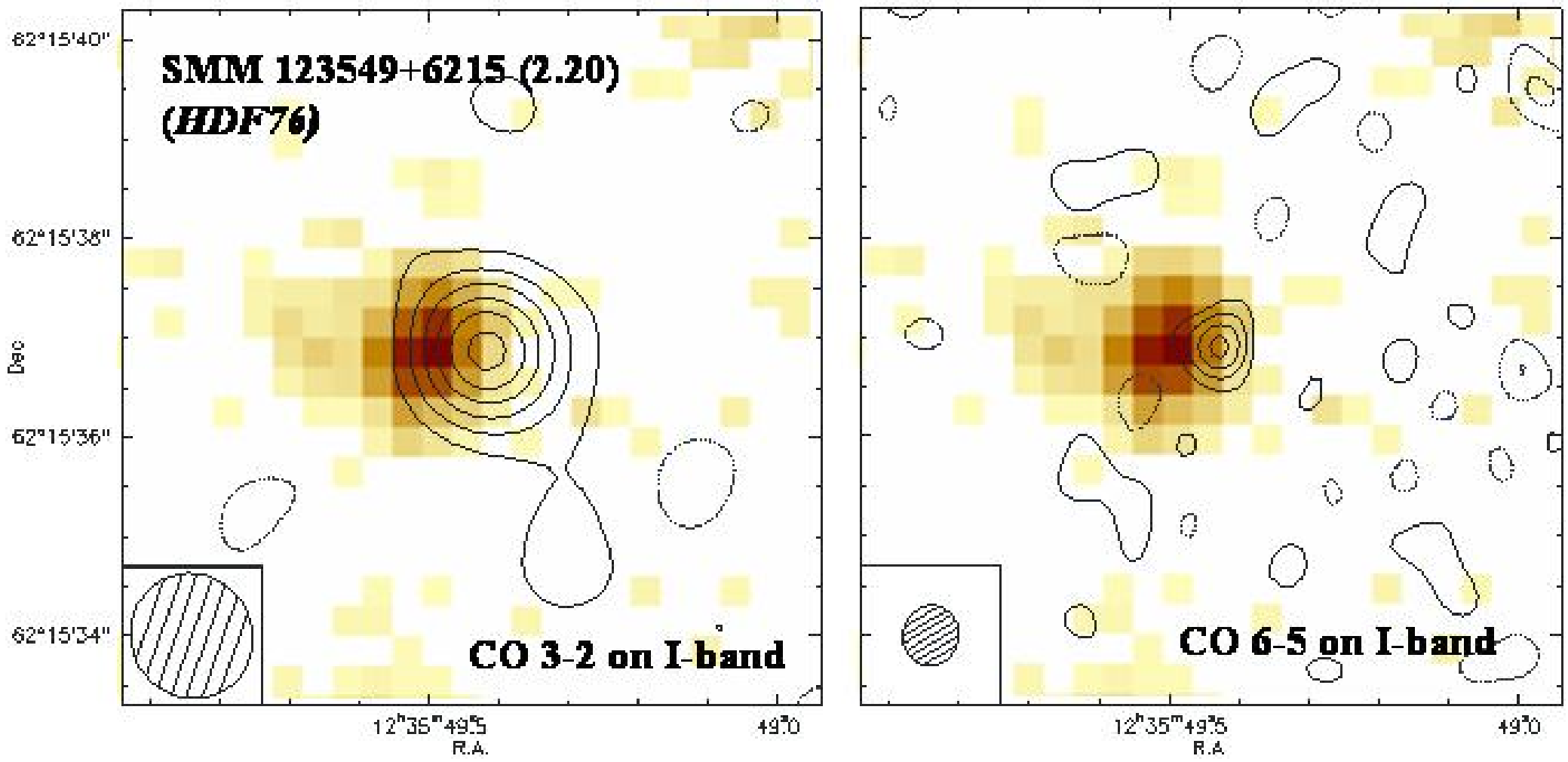

Fig 4. Integrated CO 3-2 emission (left) and integrated CO 6-5 emission (right) in SMMJ 123549+6215 (z=2.20, see Chapman et al. 2005, Swinbank et al. 2004), superposed on an I-band image. Within the relative astrometric uncertainties the CO and optical continuum emission regions are coincident. Left: The CO 3-2 map has a FWHM resolution of 1.3"x1.2" (pa=34°). The contours start at 0.4 mJy/beam, and spaced by 0.4 mJy/beam (2.5σ). Right: The CO 6-5 emission has a FWHM resolution of 0.61"x0.56" (pa = -14°). The evenly spaced contours start at 1 mJy/beam (1.5σ). A marginal 3σ (2 mJy) detection of the 1mm continuum is seen in the lower sideband.

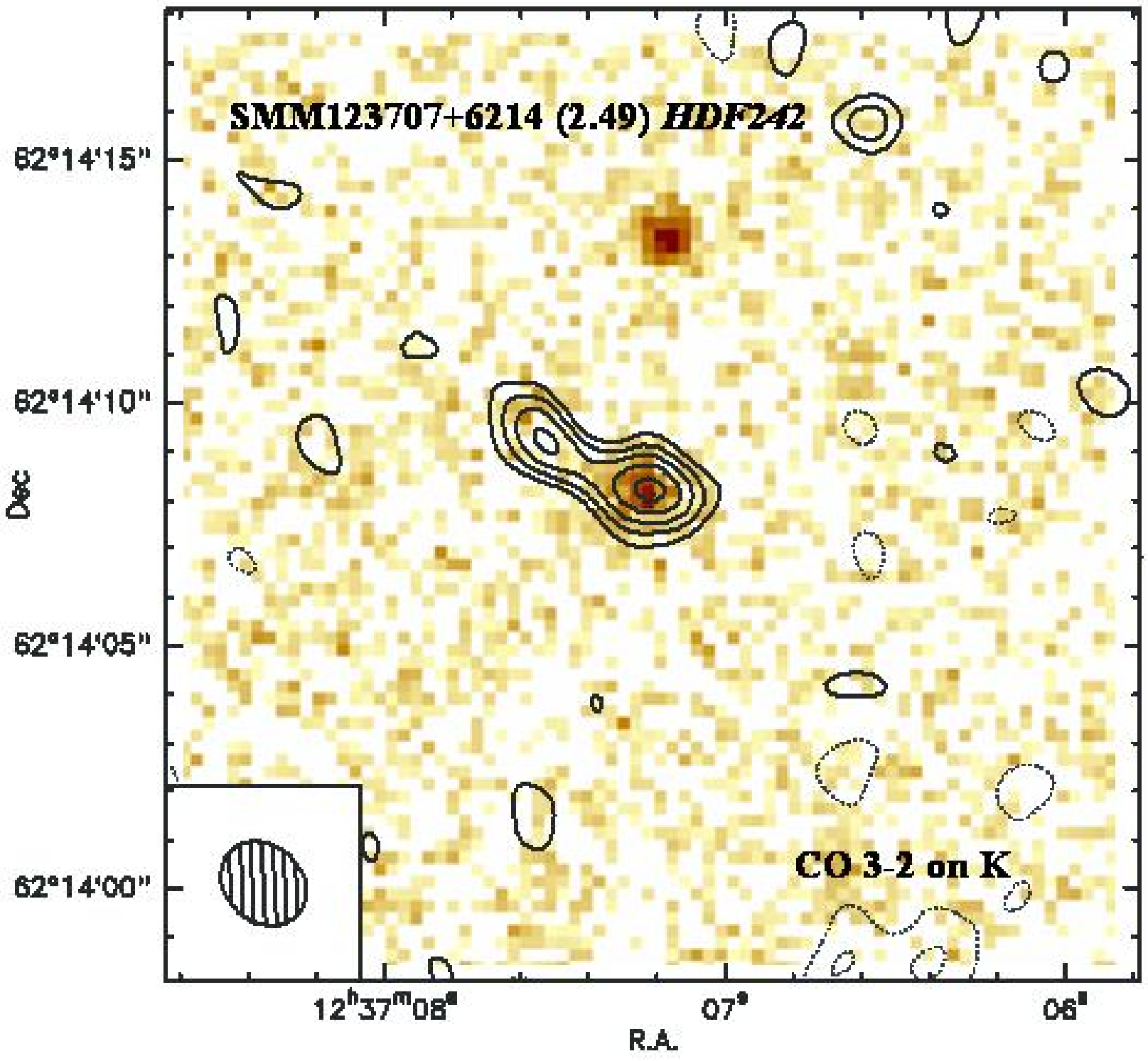

Fig.5. Integrated CO 3-2 emission of SMMJ 123707+6214 (z=2.49, see Swinbank et al. 2004; Chapman et al. 2004, 2005), superposed on a K-band image of the source (grayscale). The CO map has a FWHM resolution of 1.9”x1.6” (pa=48°). The evenly spaced contours start at 2σ in steps of 1σ (0.15 mJy/beam).

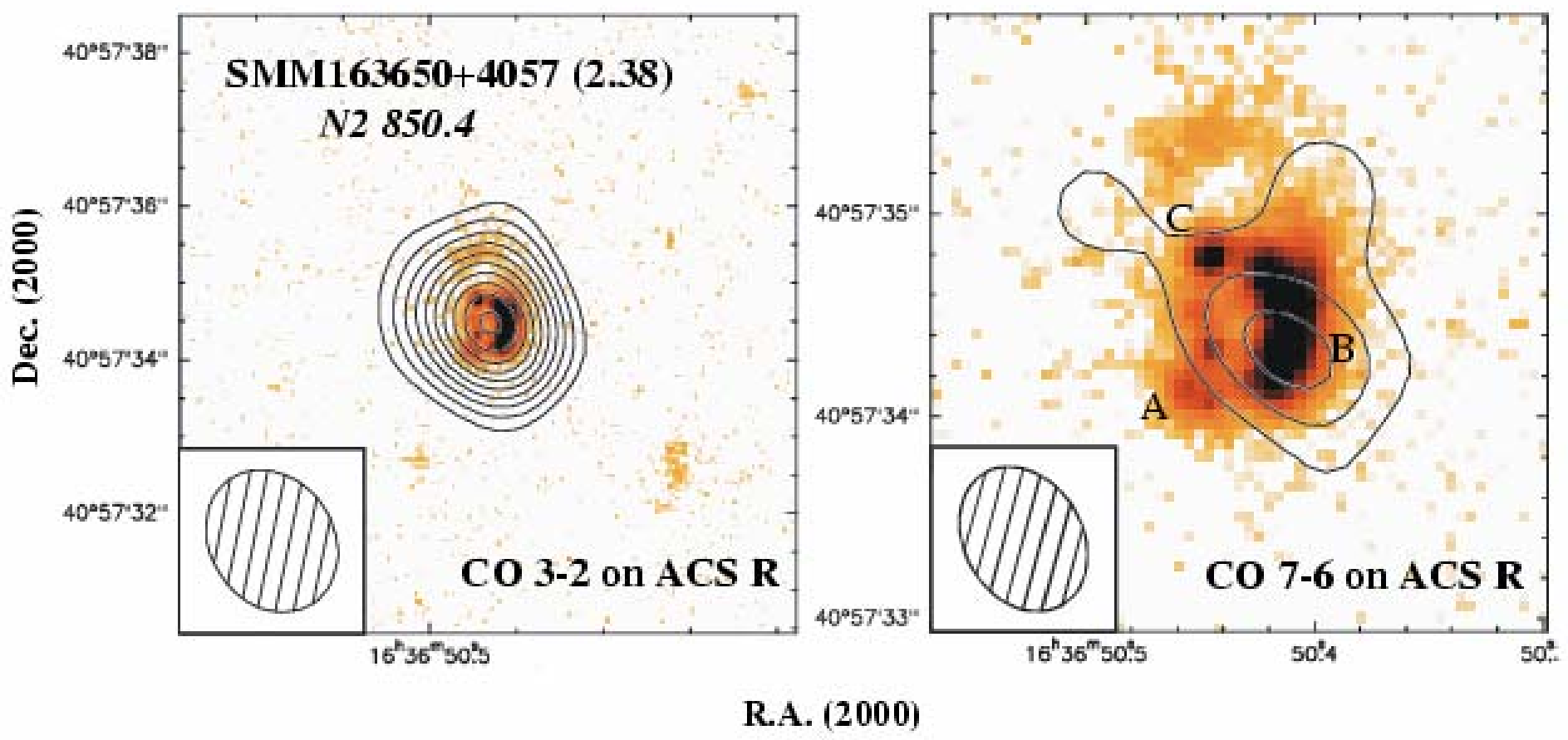

Fig.6 Left: Contour map of the integrated CO 3-2 emission of SMMJ163650+4057 (Elais N2 850.4, z=2.38; Chapman et al. 2005, Ivison et al. 2002; Neri et al. 2003), superposed on the ACS R-band image of Swinbank et al. (2005) . The synthesized beam has a FWHM of 1.9”x1.6” (pa=34°). Contours start at 0.12 mJy/beam (1.5σ), in steps of 0.12 mJy/beam. Right: Integrated 1.3mm CO 7-6 line emission at FWHM resolution of 0.74”x0.56: at p.a. 26$^{o}$. Contours are in steps of 1 times the rms noise level in the maps (0.7 mJy/beam).

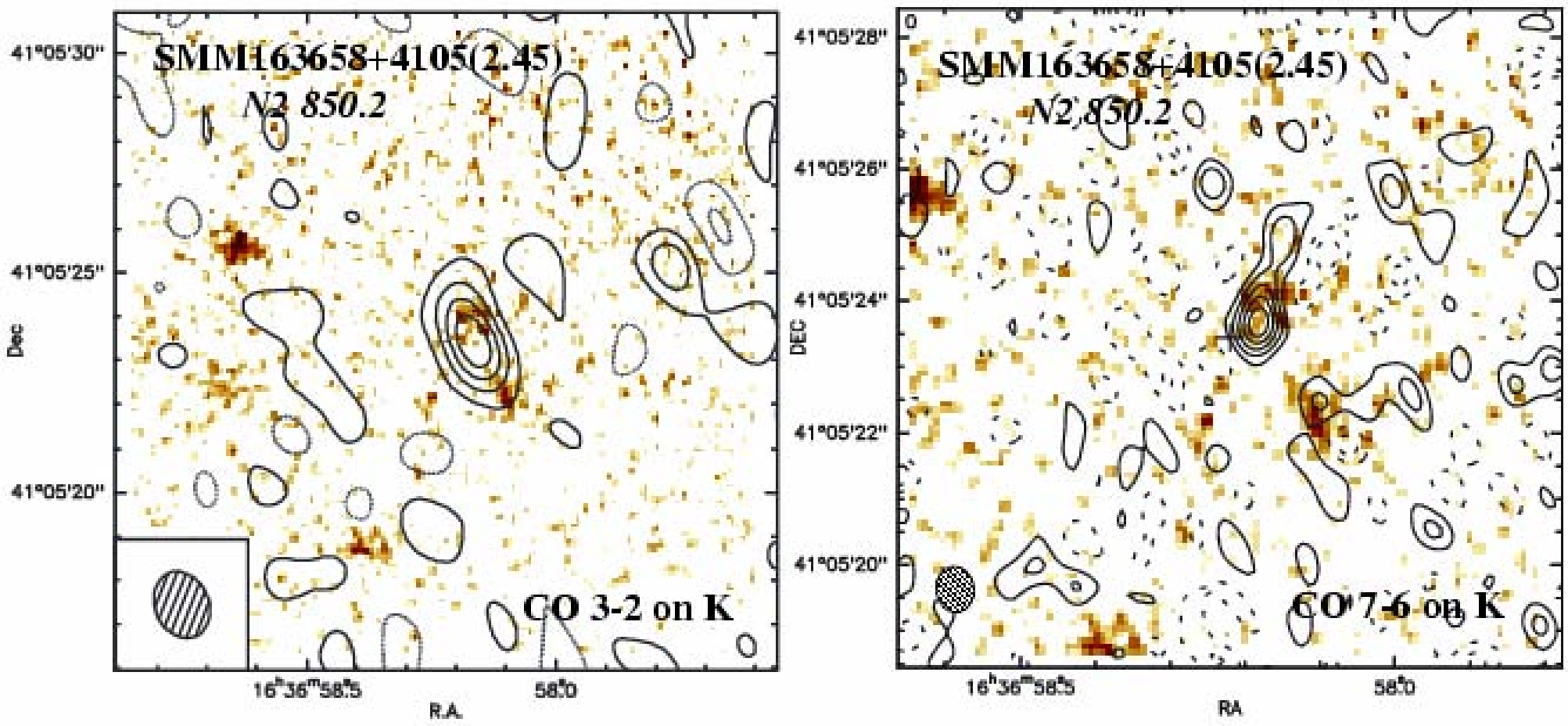

Fig.7 Left: Contour map of the integrated CO 3-2 emission of SMMJ 163658+4105 (Elais N2 850.2, z=2.45; Chapman et al. 2005; Ivison et al. 2002; Greve et al. 2005) superposed on the UKIRT K-band image of Ivison et al. (2002). The synthesized beam has a FWHM of 1.6"x1.2" (pa=23°). Contours are evenly spaced, starting at 0.2 mJy/beam (1.5σ). Right: Integrated 1.3mm CO 7-6 line emission with a synthesized beam of FWHM 0.7"x 0.56": at p.a. -169°. Contours are in steps of 1.2 times the rms noise level in the maps (1σ=0.4 mJy). The 233 GHz continuum contributes less than 1 mJy to the emission.

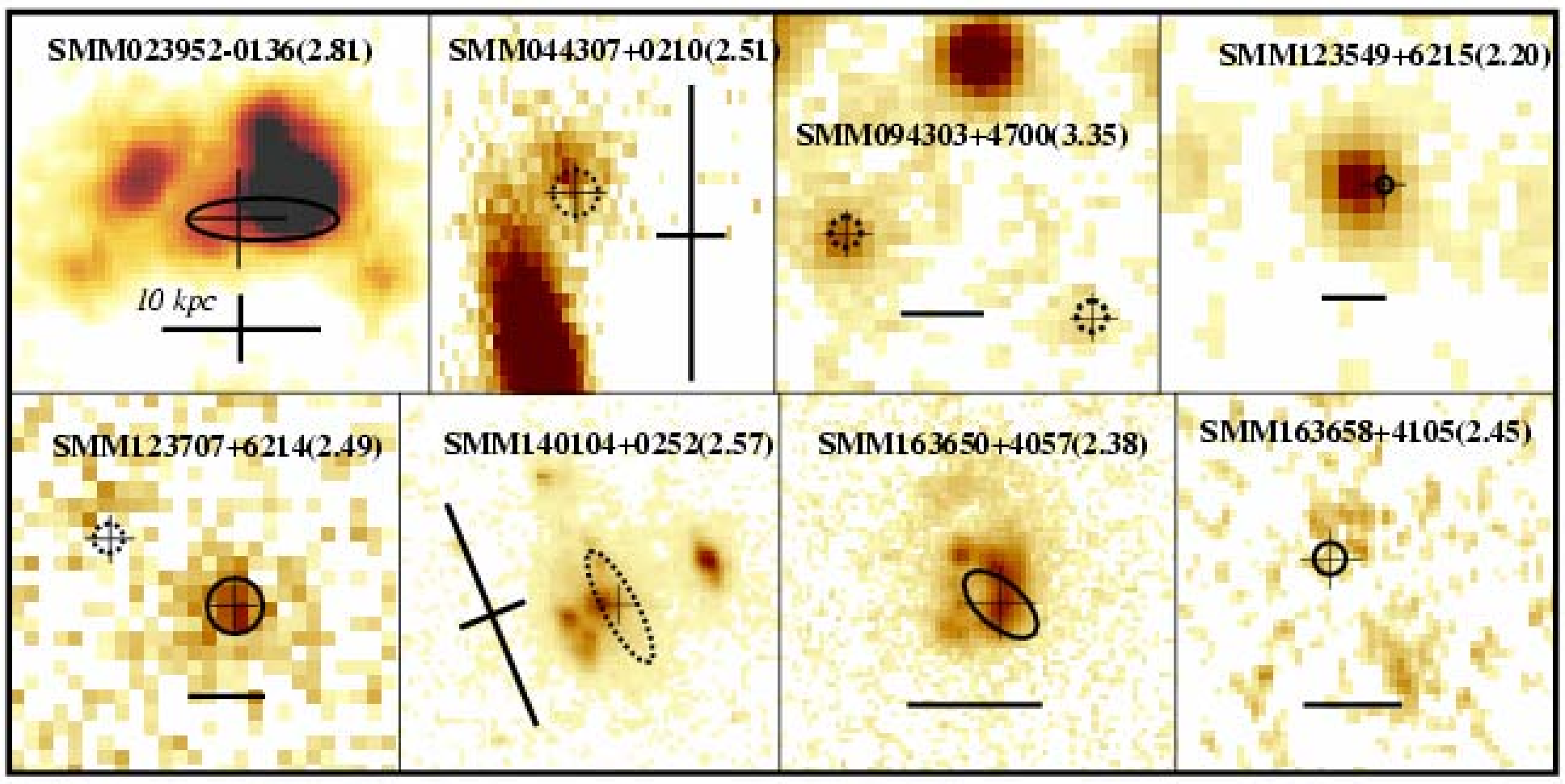

Fig.8. Composite of the mm-sizes (FWHM) of all 6 sources observed here at high resolution, plus the previously studied sources SMMJ 023952-0136 (z=2.81, Frayer et al.1998, Genzel et al. 2003) and SMMJ 140104+0252 (z=2.57, Frayer et al. 1999, Downes & Solomon 2003). In each case, we show an optical or near-infrared image, with the SMG position and its astrometric uncertainty (of optical/IR vs. mm/radio frames) shown as a thin cross. A circle/ellipse denotes our best estimate of the FWHM intrinsic mm-diameter of the source, as estimated from our line and/or continuum data. Dotted lines denote upper limits. A thick bar (or cross, for lensed sources with a preferential lensing direction) marks a length of 10 kpc in the source frame.

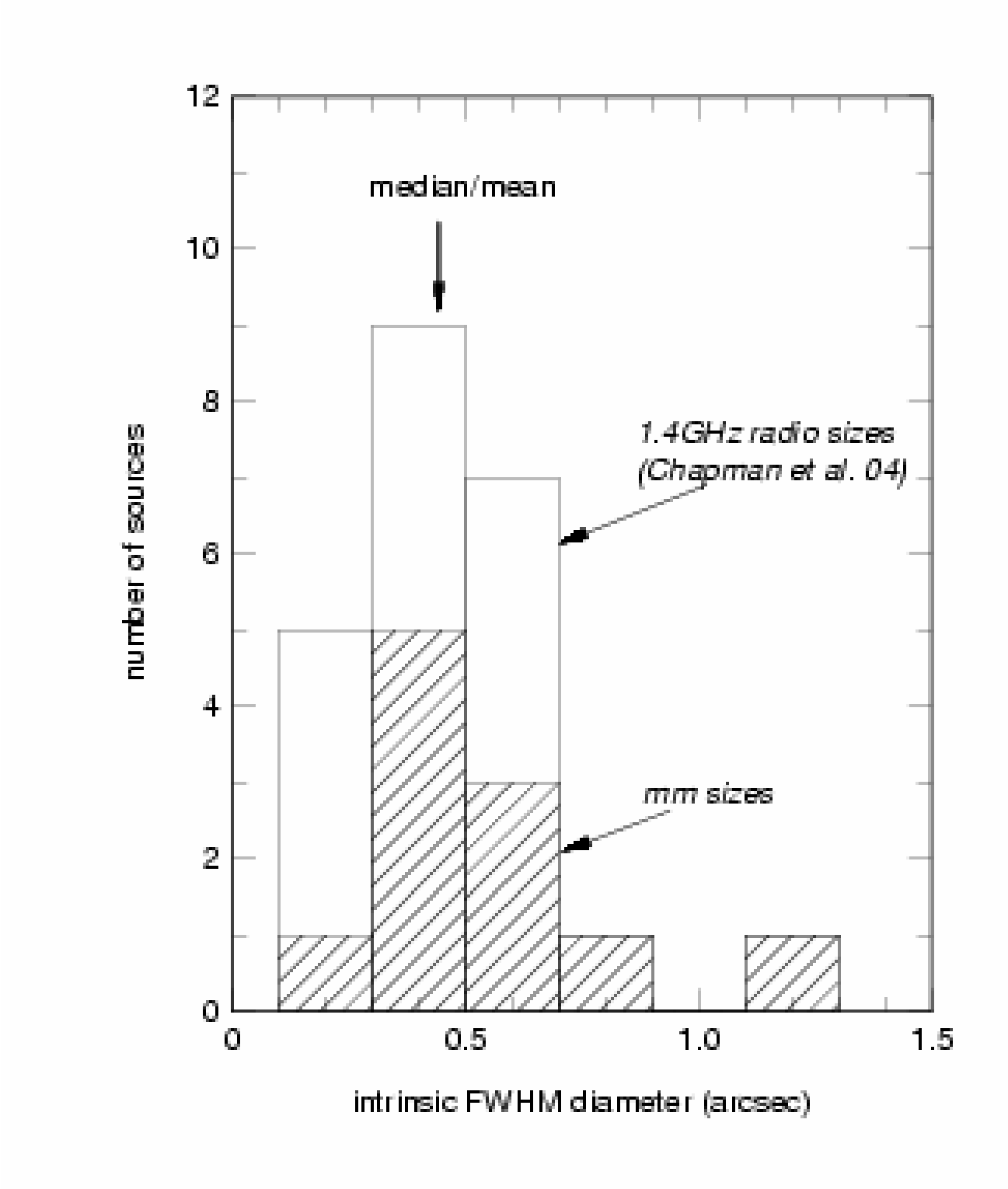

Fig. 9. Distribution of intrinsic FWHM sizes (corrected for lensing and angular averaged, and including upper limits at their nominal positions) from the high resolution mm-data in this paper, Genzel et al. (2003), Downes and Solomon (2003), and Kneib et al. (2005) (shaded), and the 1.4 GHz radio sizes from the work of Chapman et al. (2004). The mean/median of the distributions is marked by a thick arrow.

Table 1. Results from PdBI mm-interferometry

| Source | transition | redshift | R.A. (J2000) (±0.02s) | Dec. (J2000) (±0.3”) | $I_{CO}$ (Jy km/s) | $S_\nu$ (mJy) | Intrinsic FWHM size (“) | Line FWHM (km/s) |
|---|---|---|---|---|---|---|---|---|
| SMMJ044307+0210 | 1.3mm (obs) | | | | | <1.2 | | |
| | 1.3mm (predicted)* | | | | | 1.5 | | |
| | CO 3-2 | 2.509 | 04:43:07.24 | 02:10:23.8 | 1.4±0.2 | | <0.18x<0.8 | 350±40 |
| | CO 7-6 | | | | 1.0±0.3 | | --- | |
| SMMJ094303+4700 | 1.3mm(obs.) | | 09:43:04.09 | 47:00:16.2 | | 1.3±0.3 | <0.5 | --- |
| H6 | 1.3mm(predicted)* | | | | | 1.6 | | |
| | CO 4-3 | 3.346 | | | <0.4 | | | |
| SMMJ094303+4700 | 1.3mm(obs) | | 09:43:03.71 | 47:00:15.0 | | 2.2±0.3 | <0.5 | |
| H7 | 1.3mm(predicted)* | | | | | 1.0 | | |
| | CO 4-3 | 3.346 | | | 1.1±0.2 | | | 400±45 |
| SMMJ123549+6215 | 1.3mm(obs) | | | | | 2.0±0.6 | | |
| | 1.3mm(predicted)* | | | | | 1.4 | | |
| | CO 3-2 | 2.202 | 12:35:49.42 | 62:15:36.9 | 1.6±0.2 | | <0.5 | 600±50 |
| | CO 6-5 | | | | 2.3±0.4 | | 0.3±0.2 | |
| SMMJ123707+6214SW | 1.3mm(obs) | | | | | <1.4 | | |
| | 1.3(predicted)* | | | | | <2 | | |
| | CO 3-2 | 2.490 | 12:37:07.22 | 62:14:08.3 | 0.59±0.09 | | 0.9±0.3 | 430±60 |
| SMMJ123707+6214NE | 1.3mm(obs)<br>1.3mm(predicted)*<br>CO 3-2 | 2.490 | 12:37:07.57 | 62:14:09.3 | 0.32±0.09 | <1.4<br><2 | 0.5±0.2 | 430±60 |
| SMMJ163650+4057 | 1.3mm(obs)<br>1.3mm(predicted)*<br>CO 3-2<br>CO 7-6 | 2.385 | 16:36:50.41 | 40:57:34.3 | 2.3±0.3<br>6.7±0.7 | 2.6±0.5<br>2.1 | 0.8±0.2x0.4±0.3 | 710±50 |
| SMMJ163658+4105 | 1.3mm(obs)<br>1.3mm(predicted)*<br>CO 3-2<br>CO 7-6 | 2.452 | 16:36:58.17 | 41:05:23.3 | 1.8±0.2<br>3.3±0.5 | 1.5±0.5<br>2.5 | 0.4±0.2 | 800±50 |

* Predicted fluxes are from SCUBA observations as noted in the text and assuming $\beta$=1.5

Table 2. Derived Properties of Submillimeter Galaxy Population

| **Property** | **median value & uncertainty** | **comments** |
|---|---|---|
| $\langle R_{1/2} \rangle$ (kpc) | 2±0.3 | for the sources studied in this paper, plus SMMJ023952-0136 and 140104+0252 |
| $\langle v_c \rangle$ (km/s) | 392±134 | equivalent circular velocity at $R_{1/2}$ for 12 SMGs from Neri et al. (2003) and Greve et al. (2005), corrected for inclination, in the framework of orbital motion |
| $\sigma_g$ (km/s) | 100±30 | local velocity dispersion, from rotating disk fits to profile shapes of SMGs with double-peaked profiles |
| SFR ($M_\odot yr^{-1}$) | $9\pm4\text{x}10^2$ | from far-IR luminosities of SMGs (assumed $T_d$=38 K) and 1-100 $M_\odot$ Salpeter IMF such that SFR=$L(L_\odot)/(1.4\text{x}10^{10})$ |
| $\Sigma_{SFR}$ ($M_\odot yr^{-1}$ $kpc^{-2}$) | 80±20 | |
| $\langle M_{t\,½} \rangle$ ($M_\odot$) | $6\pm3\text{x}10^{10}$ | dynamical mass within $R_{1/2}$, corrected for inclination, rotating disk model, multiply by 2 for merger model |
| $f_g = \langle M_{gas}/M_t \rangle$ | 0.42±0.2 | using CO to $H_2$ conversion of Downes & Solomon (1998), including a 40% correction in mass for He, and assuming $M(HI)/M(H_2)<1$ |
| $\Sigma_t$ ($M_\odot$ $pc^{-2}$) | $5\pm1.4\text{x}10^3$ | total mass surface density from dynamics, to convert to hydrogen column density ($cm^{-2}$), multiply by $9\text{x}10^{19}$ |
| $\rho_t$ ($cm^{-3}$) | 94±35 | total mass density from dynamics, uses $2h(z)=2(\sigma_g/v_c)$ $R_{1/2}$ |
| $t_{dyn}$ (yrs) | $5\pm1.6\text{x}10^6$ | |